\documentclass[a4paper,twocolumn,accepted=2021-04-05]{quantumarticle}
\pdfoutput=1
\usepackage[utf8]{inputenc}
\usepackage[english]{babel}
\usepackage[T1]{fontenc}
\usepackage{amsmath}
\usepackage{hyperref}
\usepackage{tikz}
\usepackage{lipsum}

\usepackage{color}
\usepackage{bm}
\usepackage[linesnumbered,ruled,vlined]{algorithm2e}
\usepackage{xcolor}
\usepackage{mathtools}
\usepackage{amssymb}
\usepackage{graphicx}
\usepackage[numbers,sort&compress]{natbib}
\usepackage{braket}
\usepackage{psfrag}
\usepackage[normalem]{ulem}
\usepackage{qcircuit}
\usepackage{siunitx}

\begin{document}

\title{Grover Adaptive Search for Constrained Polynomial Binary Optimization}

\author{Austin Gilliam}
\affiliation{JPMorgan Chase}
\orcid{0000-0002-8777-0221}
\author{Stefan Woerner}
\affiliation{IBM Quantum, IBM Research -- Zurich}
\orcid{0000-0002-5945-4707}
\author{Constantin Gonciulea}
% \email{latex@quantum-journal.org}
% \homepage{http://quantum-journal.org}
\orcid{0000-0001-5870-4586}
\affiliation{JPMorgan Chase}

\maketitle

\begin{abstract}
In this paper we discuss Grover Adaptive Search (GAS) for Constrained Polynomial Binary Optimization  (CPBO) problems, and in particular, Quadratic Unconstrained Binary Optimization (QUBO) problems, as a special case.
GAS can provide a quadratic speed-up for combinatorial optimization problems compared to brute force search.
However, this requires the development of efficient oracles to represent problems and flag states that satisfy certain search criteria.
In general, this can be achieved using quantum arithmetic, however, this is expensive in terms of Toffoli gates as well as required ancilla qubits, which can be prohibitive in the near-term.
Within this work, we develop a way to construct efficient oracles to solve CPBO problems using GAS algorithms.
We demonstrate this approach and the potential speed-up for the portfolio optimization problem, i.e. a QUBO, using simulation and experimental results obtained on real quantum hardware.
However, our approach applies to higher-degree polynomial objective functions as well as constrained optimization problems.
\end{abstract}

%%%%%%%%%%%%%%%%%%%%%%%%%%%%%%%%%%%%%%%%%%%%%%%%%%%%%%%%%%%%%%%%%%%%%%%%%%%
\section{\label{sec:introduction} Introduction}
%%%%%%%%%%%%%%%%%%%%%%%%%%%%%%%%%%%%%%%%%%%%%%%%%%%%%%%%%%%%%%%%%%%%%%%%%%%
Using the laws of quantum mechanics, quantum computers offer novel solutions for resource-intensive problems. 
Quantum computers are theoretically proven to solve certain problems faster than a classical device~\cite{Grover1996, Shor1997, Lanyon2010} and are well-equipped to handle tasks such as factoring~\cite{Shor1997}, linear systems of equations~\cite{Harrow2009,bravoprieto2020variational}, Monte-Carlo simulations~\cite{Rebentrost2018Finance, Woerner2019Risk, Egger2019Credit, Stamatopoulos2019}, as well as combinatorial optimization problems~\cite{Farhi2014, Peruzzo2014, Nannicini2019, Barkoutsos2020, Braine2019}. 

A commonly-studied class of combinatorial optimization problems are \textit{Quantum Unconstrained Binary Optimization} (QUBO) problems, with applications in resource allocation, finance, machine learning, and partitioning.
There are multiple approaches to solve QUBO problems on a quantum computer, discussed in the following paragraphs.

First, quantum annealing~\cite{Childs2001, Johnson2011} is a meta-heuristic for adiabatic quantum computers.
Similar to simulate annealing, it can be used to approximate optimal solutions of QUBO problems.

Further, there exist variational quantum optimization heuristics, such as \emph{Variational Quantum Eigensolver} (VQE) and \emph{Quantum Approximate Optimization Algorithm} (QAOA)~\cite{Peruzzo2014, Farhi2014, Nannicini2019, Barkoutsos2020}.
VQE and QAOA are heuristics designed for near-term, noisy quantum computers without performance guarantees.
However, for QAOA, it is known that in the infinite depth limit, the algorithm recovers adiabatic evolution and would converge to the optimal solution.

Last, there are Grover-based~\cite{Grover1996} optimization algorithms, such as \emph{Grover Adaptive Search} (GAS)~\cite{Durr1996,Bulger2003,Baritompa2005}. 
GAS iteratively applies Grover Search to find the optimum value of an objective function, by using the best-known value as a threshold to flag all values smaller than the threshold in order to find a better solution.
The algorithmic framework comes with a quadratic speed-up, however it likely requires an error-corrected fault-tolerant quantum computer due to the depth of the resulting circuits.
One of the challenges inherent in GAS is the creation of efficient oracles.
 
In this paper, we provide a framework for automatically generating efficient oracles for solving \emph{Constrained Polynomial Binary Optimization} (CPBO)---a generalization of QUBO---with GAS.
The objective function and constraints need to be efficiently encoded, for which we use a \emph{Quantum Dictionary}~\cite{Gonciulea2019}, a pattern for representing key-value pairs as entangled quantum registers, that turns out to be efficient for polynomial functions -- in particular for quadratics representing QUBO problems. The approach relies on the addition of classical numbers to a quantum register in superposition, conditioned on the state of another quantum register. It is similar to the method used in \emph{Quantum Fourier Transform} (QFT) adders~\cite{Draper2000}. Given a boolean polynomial, the coefficient of each monomial is added to the value register conditioned on the qubits in the key register corresponding to the variables present in the monomial.

We test our algorithm on the portfolio optimization problem~\cite{Orus2019, Egger2020a}.
Multiple variants of this problem have been studied in the quantum optimization literature, ranging from convex continuous formulations~\cite{Rebentrost2018Quantum} to QUBOs~\cite{Venturelli2018a, Barkoutsos2020, Egger2020b}.
Here we investigate a QUBO formulation as well as a formulation with an inequality budget constraint, the latter not being compatible with other approaches like quantum annealing, VQE, or QAOA---as those approaches can only handle linear equality constraints through quadratic penalty terms.

The remainder of this paper is organized as follows.
Sec.~\ref{sec:grover_optimization} introduces GAS in general.
Sec.~\ref{sec:automatic_oracles} introduces QUBO problems, and shows how we can efficiently generate oracles to solve them using GAS algorithms, as well as how this approach extends to more general CPBO problems.
In Sec.~\ref{sec:test_case}, we apply the developed technique to a concrete test case -- portfolio optimization -- and demonstrate it via simulation using Qiskit~\cite{qiskit}.
Sec.~\ref{sec:conclusion} concludes this paper and discusses possible directions of future research.
 
%%%%%%%%%%%%%%%%%%%%%%%%%%%%%%%%%%%%%%%%%%%%%%%%%%%%%%%%%%%%%%%%%%%%%%%%%%%
\section{\label{sec:grover_optimization} Grover Adaptive Search}
%%%%%%%%%%%%%%%%%%%%%%%%%%%%%%%%%%%%%%%%%%%%%%%%%%%%%%%%%%%%%%%%%%%%%%%%%%%
 
Optimization problems are often solved by sequential approximation methods. 
In many cases, such methods are the only choice, but they may be computationally more efficient even when a solution to a problem can be expressed in a closed form. 
GAS works in a similar way, as it repeatedly uses Grover Search to randomly sample from all solutions better than the current one.

Grover Search is often described as a search algorithm, because it was initially formulated in the context of finding a single state of interest in a superposition of $n$-qubit quantum states.
The algorithm has been generalized to the case of multiple states of interest, in which case it is better interpreted as a sampling algorithm.
It amplifies the amplitudes of the states of interest within a larger search space, thus, increasing the probability of measuring one of the target states.

Grover Search -- the core element of GAS -- needs three ingredients:
\begin{enumerate}  
\item A state preparation operator $A$ to construct a superposition of all states in the search space. In this manuscript, $A$ is implemented by Hadamard gates $H^{\otimes n}$, i.e. it constructs the equal superposition state:
\begin{eqnarray}
H^{\otimes n} \ket{0}_n = \frac{1}{\sqrt{2^n}}\sum_{i=0}^{2^n - 1} \ket{i}_n. \label{eq:superposition}
\end{eqnarray}
\item An oracle operator $O$, that recognizes the states of interest and multiplies their amplitudes by -1. For instance, suppose $I \subset \{0, \ldots, 2^n-1\}$ denotes the set of target states and $A =H^{\otimes n}$, then
\begin{eqnarray}
OA\ket{0}_n = \frac{1}{\sqrt{2^n}}\sum_{i \notin I} \ket{i}_n - \frac{1}{\sqrt{2^n}}\sum_{i \in I} \ket{i}_n.
\end{eqnarray}
\item The Grover diffusion operator $D$, that multiplies the amplitude of the $\ket{0}_n$ state (or, equivalently, all states except $\ket{0}_n$) by -1.
\end{enumerate}

\begin{algorithm} 
  \caption{\label{algo:durr_hoyer} Grover Adaptive Search}
  \KwIn{$f: X \rightarrow \mathbb{R}$, $\lambda > 1$}
  Uniformly sample $x_1 \in X$ and set $y_1=f(x_1)$\;
  Set $k=1$\ and $i=1$\;
    \Repeat{\upshape{a termination condition is met}}{
      Randomly select the rotation count $r_i$ from the set $\{0,1,...,\lceil k-1\rceil\}$\;
      Apply Grover Search with $r_i$ iterations, using oracles $A_{y_i}$ and $O_{y_i}$.
      We denote the outputs $x$ and $y$ respectively\;
      \If{$y < y_i$}{$x_{i+1}=x$, $y_{i+1}=y$, and $k=1$}
      \Else{$x_{i+1}=x_i$, $y_{i+1}=y_i$, and $k=\lambda k$}
      $i = i+1$\;    }
\end{algorithm}

The diffusion operator has the net effect of inverting all amplitudes in the quantum state about their mean.
This causes all the amplitudes of the states of interest to be magnified, while the amplitudes of all other states are decreased.
More precisely, applying the Grover operator $G = ADA^{\dagger}O$ the right number of times to state $A\ket{0}_n$ -- i.e. evaluating $G^{r} A \ket{0}_n$ for an integer $r \geq 0$ -- will maximally amplify the amplitudes of the states of interest.
The optimal number of applications $r$ depends on the number $N=2^n$ of all states and the number $s$ of states of interest, and is equal to $\lfloor\frac{\pi}{4}\sqrt\frac{N}{s}\rfloor$. 
This implies a probability of sampling a target state of at least $1/2$, which corresponds to a quadratic speed-up compared to classical search.
Since $s$ is in general unknown, we can either use Quantum Counting algorithms~\cite{Mosca1998, Brassard2000, Suzuki2019, Aaronson2019} to find $s$, or apply a randomized strategy.

The latter is the essence of~\cite{Boyer1998}, where an algorithm for applying Grover Search for unknown $s$ is presented.
This was then used to create a minimum-finding algorithm~\cite{Durr1996}, which we refer to as GAS. In the following we outline GAS, which is formally given in Alg.~\ref{algo:durr_hoyer}.

Consider a function $f: X \rightarrow \mathbb{R}$ for $n$ binary variables, where for ease of presentation assume $X = \{0, 1\}^n$, for which we are interested in finding $\min_{x \in X} f(x)$.
The main idea of GAS is to construct $A_y$ and $O_y$ for a given threshold $y$ such that they flag all states $x \in X$ satisfying $f(x) < y$, such that we can use Grover Search to find a solution $\tilde{x}$ with a function value better than $y$.
Then we set $y = f(\tilde{x})$ and repeat until some formal termination criteria is met, e.g. based on the number of iterations, time, or progress in $y$.

While implementations of GAS vary around the specific use case~\cite{Baritompa2005, Baran2019}, the general framework still loosely follows the steps described in~\cite{Durr1996}. 
In the following, we will show how operator $A$ and oracle $O$ can be efficiently constructed for QUBO as well as CPBO problems.

%%%%%%%%%%%%%%%%%%%%%%%%%%%%%%%%%%%%%%%%%%%%%%%%%%%%%%%%%%%%%%%%%%%%%%%%%%%
\section{\label{sec:automatic_oracles}QUBO and CPBO Oracles}
%%%%%%%%%%%%%%%%%%%%%%%%%%%%%%%%%%%%%%%%%%%%%%%%%%%%%%%%%%%%%%%%%%%%%%%%%%%

A QUBO problem with $n$ binary variables is specified by a quadratic operator represented by a matrix $Q \in \mathbb{R}^{n \times n}$, vector $b \in \mathbb{R}^n$, and constant $c \in \mathbb{R}$, defined as
\begin{align} \label{eqn:qubo}
\min_{x \in \{0, 1\}^n} \left(\sum_{i,j=1}^n Q_{ij} x_i x_j + \sum_{i=1}^n b_i x_i + c\right),
\end{align}
or more compactly as $\min_{x \in \{0, 1\}^n} (x^T Q x + b^T x + c)$, i.e. $f(x) = x^T Q x + b^T x + c$.

In the following, we show how to efficiently construct GAS oracles for QUBO problems.
We will construct $A_y$ such that it prepares a $n$-qubit input register to represent the equal superposition of all $\ket{x}_n$ and a $m$-qubit output register to (approximately) represent the corresponding $\ket{f(x)-y}_m$.
Then, the oracle $O_y$ should flag the states with a negative value in the output register.
Note that in the implementation discussed, the oracle operator is actually independent of $y$, but this is not a requirement.
For clarity, we will refer to the oracle as $O$ when the oracle is independent of $y$.
More formally, we show how to construct the oracles such that 
\begin{align}
A_y \ket{0}_n\ket{0}_m &= \frac{1}{\sqrt{2^n}} \sum_{x=0}^{2^n-1} \ket{x}_n\ket{f(x)-y}_m \text{, and} \\
O \ket{x}_n\ket{z}_m &= \text{sign}(z) \ket{x}_n\ket{z}_m
\end{align}
where $\ket{x}$ is the binary encoding of the integer $x$.
Furthermore, we will show how the developed technique can be used to extend GAS to higher-degree polynomials of binary variables, as well as to constrained optimization.

\subsection{\label{subsec:construction_of_a}Construction of operator \it{A}}

To construct $A$, we will use a Quantum Dictionary, as introduced in~\cite{Gonciulea2019}, and summarize the construction in the following subsections. 

\subsubsection{\label{subsubsec:encoding-integers}Encoding a Single Integer Value}
 
Given an $m$-qubit register and an angle $\theta \in [-\pi, \pi)$, we wish to prepare a quantum state whose state vector represents a "periodic signal" equivalent to a geometric sequence of length $2^m$. 
This can be implemented using a unitary operator defined by Fig.~\ref{fig:ug-definition}.
% \begin{eqnarray}
% U_{\text{G}}(\theta) H^{\otimes m}\ket{0}_m = \frac{1}{\sqrt{2^m}}\sum_{k=0}^{2^m - 1} e^{ik\theta} \ket{k}_m.
% \end{eqnarray}
\begin{figure}[ht]
    \centering
    \mbox{
    \Qcircuit @C=1em @R=1em {
    & \lstick{H^{\otimes m}\ket{0}_m{}} &
    \gate{U_{\text{G}}(\theta)} & \qw & 
    = & & & & & & \frac{1}{\sqrt{2^m}}\sum_{k=0}^{2^m - 1} e^{ik\theta} \ket{k}_m \\ 
    }
}
\caption{\label{fig:ug-definition} Definition of unitary operator $U_G(\theta)$, where $\theta \in [-\pi, \pi)$, applied to an $m$-qubit register in equal superposition. The result is a quantum state vector that represents a geometric sequence of length $2^m$.}
\end{figure}

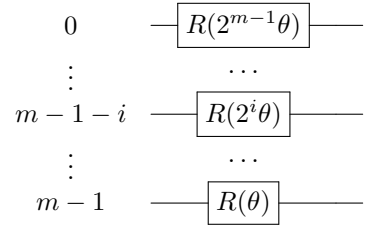
\begin{figure}[ht]
    \centering
    \mbox{
    \Qcircuit @C=1em @R=0em @!R {
    0 & {} & {} & {} & \gate{R(2^{m - 1}\theta)} & \qw & \qw  \\ 
    \vdots & & & & \ldots \\
    m-1-i & {} & {} & {} & \gate{R(2^{i}\theta)} & \qw & \qw  \\ 
    \vdots  & & & & \ldots \\
    m - 1 & {} & {} & {} & \gate{R(\theta)} & \qw & \qw   \\ 
    }
}
\caption{\label{fig:geom}
A circuit for unitary operator $U_{\text{G}}(\theta)$, where $\theta \in [-\pi, \pi)$, applied to an $m$-qubit register. $R$ denotes the phase gate, which rotates the amplitudes of states having 1 in the position corresponding to the qubit it was applied to.}
\end{figure}

The simplest implementation of $U_{\text{G}}(\theta)$ uses the phase gate $R(\theta)$ that, when applied to a qubit, rotates the phase of the amplitudes of the states having 1 in the position corresponding to that qubit. In Qiskit, this gate is the $U_1(\theta)$ operator \cite{qiskit}. 
The circuit for $U_{\text{G}}(\theta)$ is shown in Fig.~\ref{fig:geom}, and consists of applying the gate $R(2^{i}\theta)$ to the qubit $m - 1 - i$ in the $m$-qubit register prepared in the state of equal superposition in equation \eqref{eq:superposition}.

In~\cite{Gonciulea2019} an alternative way to implement $U_{\text{G}}$ was introduced by applying controlled $R_y$ rotations to an ancillary register containing the encoding of an eigenstate of $R_y$, as explained in Appendix \ref{sec:quantum_dictionary}.

Note that QFT applied to a register containing the binary encoding of a non-negative integer also creates a geometric sequence of amplitudes in that register. $U_{\text{G}}$ can be seen as a shortcut for the QFT when the encoded numbers are known classically, as we avoid multi-qubit interactions.

Given an integer $-2^{m - 1} \le k < 2^{m - 1}$, if we apply $U_{\text{G}}(\frac{2\pi}{2^{m}}k)$, followed by the inverse QFT to an $m$-qubit register prepared in the state of equal superposition in equation \eqref{eq:superposition}, we end up with ${k \pmod{2^m}}$ being encoded in the register, as shown in Fig.~\ref{fig:geom-qft-integer}. 
\begin{figure}[ht]
    \centering
    \mbox{
    \Qcircuit @C=0.8em @R=1em {
    & \lstick{H^{\otimes m}\ket{0}_m{}} &
    \gate{U_{\text{G}}(\frac{2\pi}{2^{m}}k)}  & \gate{QFT^\dag} & \qw & 
    = & & & & &  \ket{k \pmod{2^m}}_m{} \\ 
    }
}
\caption{\label{fig:geom-qft-integer}The geometric sequence encoding of an integer $-2^{m - 1} \le k < 2^{m - 1}$, applied to a register of $m$ qubits in equal superposition.}
\end{figure}

This representation is called the binary Two's Complement of $k$, which just adds $2^m$ to negative values $k$, similar to the way we can represent negative angles with their complement, e.g. equating $-\pi/4$ with $7\pi/4$. The reason this representation occurs naturally in this context is due to the fact that rotation composition behaves like modular arithmetic.
The same method can be used to encode non-integers, as discussed in Appendix~\ref{sec:handling-non-integers}.
 
\subsubsection{\label{subsubsec:encoding-polynomials}Encoding a Superposition of Polynomial Values}
 
\begin{figure}[t]
    \centering
    \mbox{
    \Qcircuit @C=1em @R=0em @!R {
    & \lstick{\ket{x_0}{}} & \qw & \qw & \qw & \qw  \\
    & \lstick{\ket{x_1}{}} & \qw & \ctrl{3} & \qw & \qw \\
    & \lstick{\ket{x_2}{}} & \qw & \qw & \qw & \qw  \\
    & \lstick{\ket{x_3}{}} & \qw & \ctrl{1} & \qw & \qw  \\
    & \lstick{\ket{z}_m{}} & \qw & \gate{U_G(\frac{2\pi}{2^m}a_{13})} & \qw & \qw  \\ 
    }
}
\caption{\label{fig:controlled-u-example}An example of $C^{\{1,3\}}(U)$ with $n=4$ input qubits and $m$ output qubits.}
\end{figure}
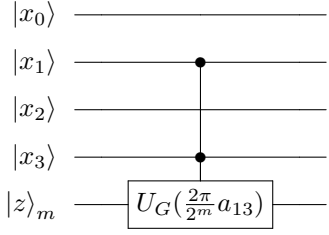

Next we will discuss the application of $U_{\text{G}}(\theta)$ to a register $\ket{z}_m$ representing the values of a function, controlled on a register $\ket{x}_n$ representing the inputs of the same function. In general, the  application of a unitary operator $U$ to $\ket{z}_m$ controlled by a set $J \subseteq \{0, \ldots, n - 1\}$ of qubits of $\ket{x}_n$ can be expressed as
\begin{eqnarray}
C^J(U)\ket{x}\ket{z} = \ket{x}U^{\prod\limits_{j \in J} x_j}\ket{z},
\end{eqnarray}
and an example is shown in Fig.~\ref{fig:controlled-u-example}.
 
The general form of a polynomial of $n$ variables is: 
\begin{eqnarray}
P(x) = \sum\limits_{J \subseteq \{0, \ldots, n - 1\}}a_J \prod\limits_{j \in J} x_j.
\end{eqnarray}
Each subset $J \subseteq \{0, \ldots, n - 1\}$ has a corresponding monomial $\prod\limits_{j \in J} x_j$. The free term is represented by  $a_{\emptyset}$.

\subsubsection{\label{subsubsec:def-operator-a}Construction of Operator A}
 
\begin{figure}[ht]
    \centering
    \mbox{
    \Qcircuit @C=1em @R=0em @!R {
    & \lstick{\ket{x}{}_n} & \gate{H} & \ctrl{1} & \ctrl{1} & \ctrl{1} & \qw & \qw  \\
    & \lstick{\ket{z}{}_m} & \gate{H} & \gate{\text{\ldots}} &  \gate{U_{\text{G}}(\frac{2\pi}{2^{m}}a_J)} &
    \gate{\text{\ldots}} & \gate{QFT^\dag} & \qw  
    }
}
\caption{\label{fig:operatorA}The circuit for state preparation operator $A$, applied to input register $\ket{x}_n$ and output register $\ket{z}_m$. Starting in a state of equal superposition, the operator employs several controlled applications of the unitary operator $U_G$, whose angle parameter corresponds to a non-zero coefficient $a_J$, where $J$ is a subset of $\{0, \ldots, n - 1\}$. The single application of the inverse Quantum Fourier Transform ($QFT^{\dagger}$) at the end of the circuit decodes the periodic signal encoded by $U_G$, resulting in a superposition of key-value pairs.}
\end{figure}
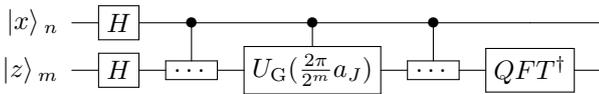

Now we are ready to define the operator $A$, as shown in Fig.~\ref{fig:operatorA}.
It consists of applying a controlled geometric sequence transformation $C^J(U_{\text{G}}(\frac{2\pi}{2^{m}}a_J))$ for each subset $J \subseteq \{0, \ldots, n - 1\}$ with a non-zero coefficient $a_J$, followed by a single application of the inverse QFT.
 
\begin{figure}[ht]
    \centering
    \mbox{
    \Qcircuit @C=1em @R=0em @!R {
    & \lstick{\ket{x_0}{}} & \qw & \qw      & \qw & \qw         & \qw       & \qw       & \qw  \\
    & \lstick{\ket{x_1}{}} & \qw & \ctrl{3} & \qw & \ctrl{4}    & \qw       & \ctrl{5}  & \qw \\
    & \lstick{\ket{x_2}{}} & \qw & \qw      & \qw & \qw         & \qw       & \qw       & \qw  \\
    & \lstick{\ket{x_3}{}} & \qw & \ctrl{1} & \qw & \ctrl{2}    & \qw       & \ctrl{3}  & \qw \\
    & \lstick{\ket{z_0}{}} & \qw & \gate{R(2\pi)} & \qw & \qw         & \qw       & \qw       & \qw  \\ 
    & \lstick{\ket{z_1}{}} & \qw & \qw      & \qw & \gate{R(\pi)}    & \qw       & \qw       & \qw \\ 
    & \lstick{\ket{z_2}{}} & \qw & \qw      & \qw & \qw         & \qw       &\gate{R(\pi/2)}   & \qw \\ 
    }
}
\caption{\label{fig:monomial-example}An example of encoding the monomial $2x_1x_3$, with input register $\ket{x}_4$ and output register $\ket{z}_3$. $R$ denotes the phase gate, which rotates the amplitudes of states having 1 in the position corresponding to the qubit it was applied to.}
\end{figure}
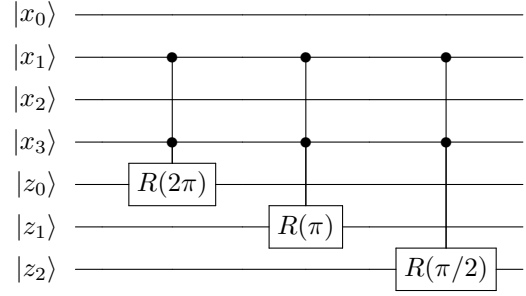

Note that QUBO polynomials have only monomials of degree less than or equal to $2$, i.e. $|J| \le 2$, so we only need to control on single and pairs of qubits.
An example of encoding a single monomial is shown in Fig.~\ref{fig:monomial-example}. 
 
\begin{table}
\centering
\begin{tabular}{c|c}
\textbf{Gate} & \textbf{Count} \\
\hline
$H$ & $m$ \\
$R$ & $m$ \\
$1$-controlled $R$ & $m \cdot n$ \\
$2$-controlled $R$ & $m \cdot n (n - 1) / 2$ \\
Inverse $QFT$ on $m$ qubits & $1$ \\
\end{tabular}
\caption{\label{tab:gate-analysis}Number of required gates to realize $A$ in terms of number of input qubits $n$ and output qubits $m$ assuming a QUBO with dense $Q$ and $b$ and non-zero offset $c$. The depth scales as $1/m$-times the number of gates since they can be mostly applied in parallel. Note that a $QFT$ on $m$ qubits can be implemented using $O(mlog(m))$ gates and thus will not dominate the overall circuit complexity~\cite{Nam2020}}
\end{table}
 
The operator $A$ prepares a state where the $\ket{x}_n$ register holds all $2^n$ inputs in equal superposition, entangled with the corresponding values $P(x)$ encoded in the $\ket{z}_m$ register:
\begin{eqnarray}
A \ket{0}_n\ket{0}_m &=& \frac{1}{\sqrt{2^n}} \sum_{x=0}^{2^n-1} \ket{x}_n\ket{P(x)}_m,
\end{eqnarray} 
where we are assuming that an appropriate size $m$ for the value register is known.
 
The desired $A_y$ operator is obtained by adding the $-y$ constant to the free term of the original polynomial.
This construction works for polynomials of arbitrary degree.

A detailed summary of the number of gates required to construct $A$ for a QUBO is given in Table~\ref{tab:gate-analysis}.
In general, the number of gates scales as the number of monomials in the polynomial to be loaded.
This matches the scaling of the description of the problem under the assumption that the input weights and output are represented using roughly the same number of bits, thus, the asymptotic scaling of our approach is optimal.

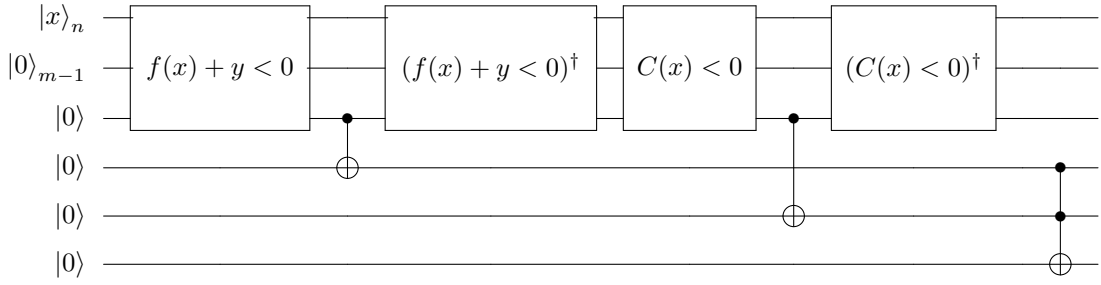
\begin{figure*}[ht]
    \centering
    \mbox{
    \Qcircuit @C=1em @R=1em {
    & \lstick{\ket{x}_n{}}      & \multigate{2}{f(x)+y<0}   & \qw       & \multigate{2}{(f(x)+y<0)^{\dagger}} & \multigate{2}{C(x)<0} & \qw      & \multigate{2}{(C(x)<0)^{\dagger}} & \qw & \qw & \qw \\ % index
    & \lstick{\ket{0}_{m-1}{}}  & \ghost{f(x)+y<0}          & \qw       & \ghost{(Q(x)+y<0)^{\dagger}}        & \ghost{C(x)<0}        & \qw      & \ghost{(C(x)<0)^{\dagger}}        & \qw & \qw & \qw \\ % value (m-1)
    & \lstick{\ket{0}{}}        & \ghost{Q(x)+y<0}          & \ctrl{1}  & \ghost{(Q(x)+y<0)^{\dagger}}        & \ghost{C(x)<0}        & \ctrl{2} & \ghost{(C(x)<0)^{\dagger}}        & \qw & \qw & \qw \\ % value sign bit
    & \lstick{\ket{0}{}}        & \qw                       & \targ     & \qw                          & \qw                   & \qw      & \qw                        & \qw & \ctrl{1} & \qw \\  % value sign bit copy
    & \lstick{\ket{0}{}}        & \qw                       & \qw       & \qw                          & \qw                   & \targ    & \qw                        & \qw & \ctrl{1} & \qw \\ % constraint bits (top)
    & \lstick{\ket{0}{}}        & \qw                       & \qw       & \qw                          & \qw                   & \qw      & \qw                        & \qw & \targ    & \qw  \\ % oracle control bit
    }
}
\caption{\label{fig:constrained-optimization-example}A constrained optimization circuit, which encodes the functions corresponding to a given set of constraints into an $m$-qubit register, and flips a related indicator qubit for each condition that is satisfied. In this example, given a function $f: X \rightarrow \mathbb{R}$ of $n$ binary variables, the global indicator qubit (shown at the bottom of the circuit) is set to $\ket{1}$ if and only if $f(x)+y$---where $y$ is the threshold parameter from GAS---and the cost of the input (denoted $C(x)$) are both less than $0$. Note that the encoding method is the same as described in Sec.~\ref{subsec:construction_of_a}, meaning that the binary representation of the values is in Two's Complement, and thus we only need to control on a single qubit (the most significant bit) to determine if a value is negative. There is a trade-off between reusing the $m$ qubits and reversing the computation (shown here), or adding additional value qubits for each constraint, allowing one to skip the uncompute. In either case, we can replace the oracle that flags "good states" by a multi-controlled $Z$ gate, with a number of controls equal to one more than the number of constraints. Here, we add one qubit and use a Toffoli gate to get back to the setting introduced before.}
\end{figure*} 

An alternative approach to construct $A$ would be to use quantum arithmetic.
However, even uncontrolled in-place addition of a classically given $m$-bit number to a $m$-bit quantum register requires $2m$ qubits and $2m-1$ Toffoli gates~\cite{Cuccaro2004, Draper2004, Gidney2018}.
This then would have to be done $O(n^2)$ times in a 2-controlled way, which leads to significantly larger circuits and $m$ more qubits than required by our approach, and also requires explicit encoding of negative integers.
 
\subsection{\label{subsec:construction_of_o} Construction of oracle \it{O}}

At each step of the algorithm, we are adding a constant to the polynomial, and searching for remaining negative values. This means the oracle just needs to recognize negative integers.
Since values are represented in Two's Complement, where the most significant (left-most) bit designates the sign of the number, a single qubit in the value register can be used to recognize negative integers.
The typical oracle that multiplies target amplitudes by $-1$ can be applied.
Note that the oracle stays unchanged between iterations, because we add a constant to the polynomial, which may lead to overflow in the value register. In order to avoid that, we may need to increase the number of qubits in the value register by 1.
 
Alternatively, threshold-based oracles (which are potentially more expensive) can be used, that will reduce the search space by filtering the numbers above a given threshold.

\subsection{\label{subsec:constrained_optimization} Constrained Optimization}
 
It is common for optimization problems to impose additional constraints---e.g. the total number or cost of assets in a portfolio may be subjected to an upper bound.
Such constraints translate into further reductions of the search space based on the key register.
For example, the number of assets in a portfolio corresponds to the number of $1$s in the binary representation of the input of the objective function, called the Hamming weight. 
 
It is straightforward to use the GAS oracles introduced in Sec.~\ref{subsec:construction_of_a} and \ref{subsec:construction_of_o} to take into account polynomial equality and inequality constraints on the key register.
Therefore, we can add additional registers to evaluate other polynomials.
Whether an inequality constraint is satisfied or not can again be mapped to the sign qubit by applying an appropriate shift to the polynomial.
Equality constraints are a bit more expensive, as they require the detection of a particular state, which essentially has the same complexity as the Grover diffusion operator $D$.

This leads to a set of qubits flagging target states: one qubit identifying the states that correspond to objective values below the current threshold, and one qubit for each constraint.
Applying a logical AND-operation to all of them essentially acts as an intersection of the individually flagged states and allows to construct oracles for CPBO.
An example is shown in Fig.~\ref{fig:constrained-optimization-example}, which we demonstrate in Sec.~\ref{subsec:portfolo_optimization}.
 
%%%%%%%%%%%%%%%%%%%%%%%%%%%%%%%%%%%%%%%%%%%%%%%%%%%%%%%%%%%%%%%%%%%%%%%%%%%
\section{\label{sec:test_case}Test Cases}
%%%%%%%%%%%%%%%%%%%%%%%%%%%%%%%%%%%%%%%%%%%%%%%%%%%%%%%%%%%%%%%%%%%%%%%%%%%
In the remainder of this paper we demonstrate the proposed technique on the portfolio optimization problem, and then show a simple example on quantum hardware.

\subsection{\label{subsec:portfolo_optimization} Portfolio Optimization}

Suppose an investment universe consisting of $n$ assets, denoted by $i = 1, \ldots, n$, their corresponding expected returns $\mu \in \mathbb{R}^n$ and covariance matrix $\Sigma \in \mathbb{R}^{n \times n}$.
Furthermore, we consider a given risk factor $q \geq 0$, which determines the considered risk appetite.
The resulting objective function is
\begin{eqnarray}
\min_{x \in \{0, 1\}^n} \left(q x^T \Sigma x - \mu^T x\right).
\end{eqnarray}
In other words, we want to minimize the weighted variance minus the expected portfolio return. Setting $q = 0$ implies a risk neutral investor, while increasing $q$ increases its risk averseness.

In the presented form, portfolio optimization is a QUBO problem.
We can extend it by imposing a budget constraint of the form 
\begin{eqnarray}
\sum_{i=1}^n x_i &=& B, \label{eq:budget_equality}
\end{eqnarray}
where $B \in \{0, \ldots, n\}$ denotes the number of assets to be chosen.

In general, equality constraints can be recast as penalty terms 
\begin{eqnarray}
\lambda \left( \sum_{i=1}^n x_i - B \right)^2,
\end{eqnarray}
and added to the objective (since we minimize),
where $\lambda > 0$ is a large number to enforce the constraint to be satisfied.
This results in a quadratic term that will again lead to a QUBO problem.

However, with the methodology introduced in Sec.~\ref{subsec:constrained_optimization}, we can also model more complex constraints, e.g. a budget inequality constraint of the form
\begin{eqnarray}
\sum_{i=1}^n c_i x_i &\leq& B, \label{eq:budget_inequality}
\end{eqnarray}
where $c_i \in \mathbb{R}$ denote the asset prices, which does usually make more sense in practice than (\ref{eq:budget_equality}).
In the following we consider two examples, one with an objection function that we want to minimize, and then the same problem with added constraints.

\subsubsection{Finding a Minimum Value} %%%%%%%%

Consider a portfolio of $n=3$ assets, risk factor $q = 0.5$, and returns described by:
\small
\begin{eqnarray}
\mu = \left(  
\begin{array}{c}
1 \\
-2 \\
3
\end{array}
\right)
& \quad \text{and} \quad &
\Sigma = \left(
\begin{array}{ccc}
2 & 0 & -4 \\
0 & 4 & -2 \\
-4 & -2 & 10
\end{array}
\right)
\end{eqnarray}
\normalsize
which leads to the formulation
\begin{eqnarray}
\min_{x \in \{0, 1\}^3} \left(-2x_1x_3 - x_2x_3 - 1x_1 + 2x_2 - 3x_3\right). \label{eq:port-opt-objective-func}
\end{eqnarray}

The objective function with added constant $-y$ (where $y$ is the current threshold) has an associated $A_y$ operator and the oracle $O$ recognizes negative values, as introduced in Sec.~\ref{sec:automatic_oracles}.
To perform the experiment we need $7$ qubits split into two registers, $n=3$ input qubits and $4$ output qubits. While we only need $3$ qubits in the output register, we add $1$ extra to accommodate for the threshold shift.
We set the initial threshold $y_1=0$, and stop searching if an improvement has not been found in three consecutive iterations of the algorithm.

For each iteration of GAS, we determine the number of applications of the Grover iterate as defined in Alg.~\ref{algo:durr_hoyer}. 
We apply $A_y$ for the current threshold, and then apply the Grover iterate $G_y = A_yDA_y^{\dagger}O$ for the predetermined number of applications. 
If the measured value is less than $y$, we update the threshold. 
This process repeats until we have seen no improvement for three consecutive iterations. 

\begin{figure}[t]
\centering
\includegraphics[width=8cm]{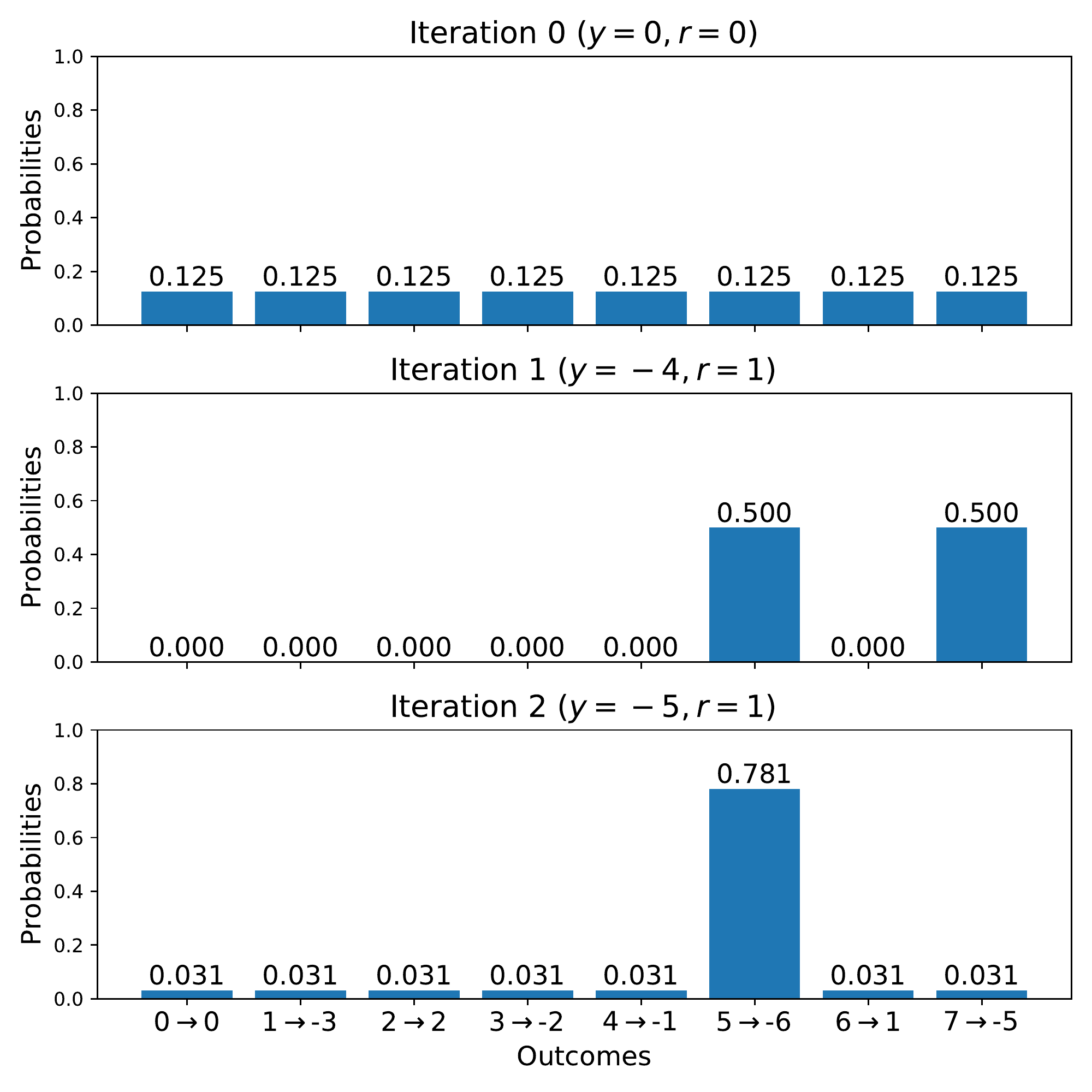}
\caption{\label{fig:portfolio-optimization-example}The output probabilities of GAS for three iterations, searching for a minimum value of a function (Eq.~\ref{eq:port-opt-objective-func}), each with thresholds $y$ and $r$ applications of the Grover operator.}
\end{figure}

Classically, we can determine the original minimum value by keeping track of the total shift, or by calculating the value of the objective function for the minimum key.
The results of this simulated experiment are shown in Fig.~\ref{fig:portfolio-optimization-example}. 

\subsubsection{\label{subsubsec:additional-constraints}Additional Constraints} %%%%%%%%

\begin{figure}[ht]
\centering
\includegraphics[width=8cm]{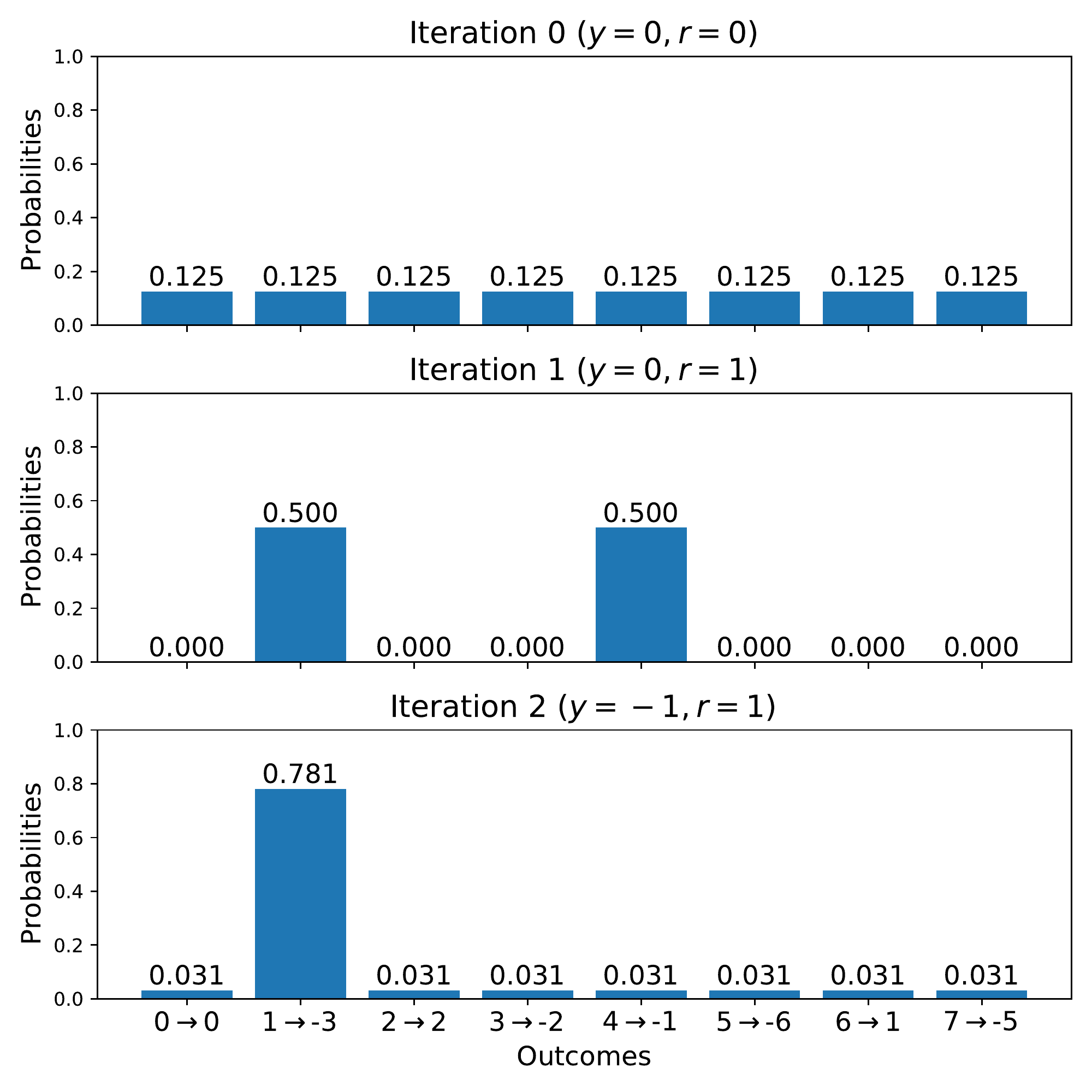}
\caption{\label{fig:portfolio-optimization-example-constraint}The output probabilities of three iterations of GAS, with respective thresholds $y$ and $r$ applications of the Grover operator. We want to find a minimum value of Eq.~\ref{eq:port-opt-objective-func}, with the additional constraint that the binary representation of the corresponding key has a Hamming weight less than $2$.}
\end{figure}

We can impose a budget inequality constraint of the form discussed in Eq.~\ref{eq:budget_inequality} to the previous problem, where $B < 2$ and the price of each asset is $1$.
As shown in Sec.~\ref{subsec:constrained_optimization}, we can implement this constraint by adding an additional register to the existing quantum circuit, and then encode the Hamming weight of the binary representation of the keys in that register.
Note that we only need to control on the most significant qubit of the constraint register to determine if $B < 2$.
Otherwise, the procedure for applying GAS is the same as in the original problem.
The results of this simulated experiment are shown in Fig.~\ref{fig:portfolio-optimization-example-constraint}.

%%%%%%%%%%%%%%%%%%%%%%%%%%%%%%%%%%%%%%%%%%%%%%%%%
\subsection{\label{subsec:real-hardware}Trials on Real Hardware}
%%%%%%%%%%%%%%%%%%%%%%%%%%%%%%%%%%%%%%%%%%%%%%%%%
The experiments discussed in this section were run on the IBM ibmq\_toronto device, with Quantum Volume 32~\cite{QuantumVolume}. The configuration details for ibmq\_toronto at the time of our experiments is given in Appendix~\ref{sec:hardware-config}.
Readout error-mitigation techniques were applied to the results of each circuit~\cite{qiskit, Dewes2012}.

Let us consider a simple example of a polynomial minimization which can be run on current quantum hardware:
\begin{eqnarray}
\min_{x \in \{0, 1\}^2} (-2 + x_1 + x_2).
\end{eqnarray}

As in the previous subsection, we set the initial threshold $y_1=0$, and stop searching after three iterations with no improvement.
Note that due to the probabilistic nature of the algorithm there is a non-zero probability that an invalid key-value pair will be measured.
In addition, the noise inherent in the present era of quantum hardware further impacts the results and increases the probability of wrong results.
We repeat the computation several times, and take the outcome with the maximum probability as the measured result.
As long as the noise is not too strong, this still achieves a good key-value pair.

\begin{figure}[t]
\centering
\includegraphics[width=8cm]{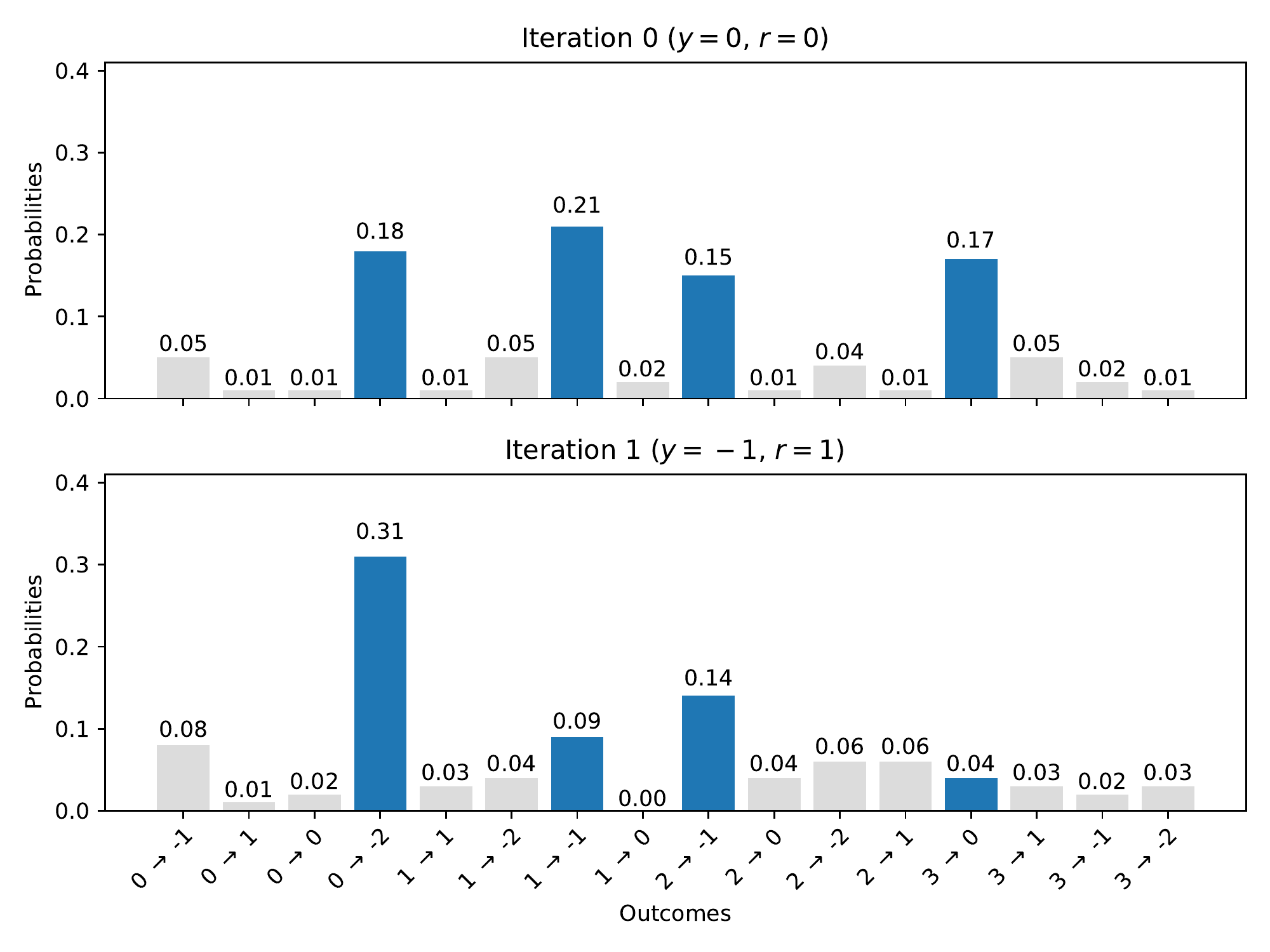}
\caption{\label{fig:real-hardware-example}The output probabilities of GAS for two iterations run on real quantum hardware, with respective thresholds $y$ and $r$ applications of the Grover operator. The valid measurement outcomes are shown in blue, while the invalid measurement outcomes are shown in grey.}
\end{figure}

The results of the experiment are shown in Fig.~\ref{fig:real-hardware-example}.
Note that in this example, the valid measurement outcomes are $0 \rightarrow -2$, $1 \rightarrow -1$, $2 \rightarrow -1$, and $3 \rightarrow 0$.
In the first iteration the four outcomes are shown in an approximately-equal superposition, and sampled randomly (we do not apply the Grover iterate).
We measure $1 \rightarrow -1$, and thus we update the threshold to $y_2=-1$ and the number of iterations to $r=1$.
In the second iteration of Grover Adaptive Search the results of the amplification are shown, where $0 \rightarrow -2$ (the minimum) has the highest probability.

%%%%%%%%%%%%%%%%%%%%%%%%%%%%%%%%%%%%%%%%%%%%%%%%%%%%%%%%%%%%%%%%%%%%%%%%%%%
\section{\label{sec:conclusion} Conclusion}
%%%%%%%%%%%%%%%%%%%%%%%%%%%%%%%%%%%%%%%%%%%%%%%%%%%%%%%%%%%%%%%%%%%%%%%%%%%
 
In this paper we introduced an efficient way to implement the oracles required for solving \emph{Constrained Polynomial Binary Optimization } problems using \emph{Grover Adaptive Search}.
This problem class is very general and contains for instance QUBO problems.
Our approach significantly reduces the number of gates required compared to standard quantum arithmetic approaches, i.e. it lowers the requirements to apply GAS on real quantum hardware for practically relevant problems.
We demonstrated our algorithm on the portfolio optimization problem, i.e. a QUBO, where we could reliably find the optimal solution, and on real quantum hardware.
Within this manuscript we focused mainly on problems with integer coefficients.
The handling of non-integers is discussed in Appendix~\ref{sec:handling-non-integers}.

\begin{figure*}[ht]
\centering
\includegraphics[width=17cm]{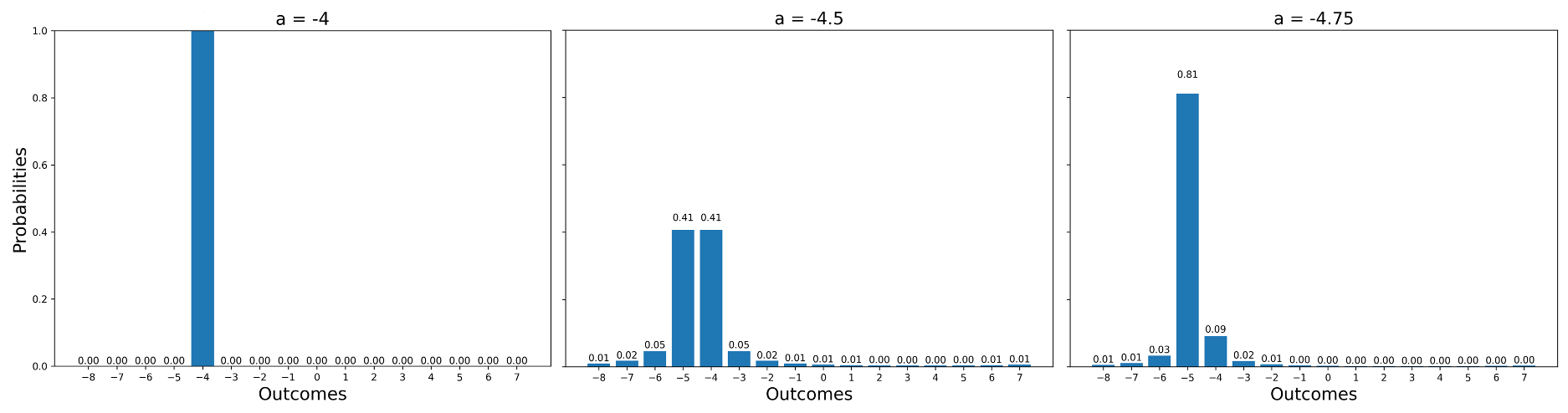}
\caption{\label{fig:fejer-example}An example of an integer encoding (left), accompanied by two examples of a non-integer encoding (middle and right). Note that in the integer case only one outcome is possible, whereas the encoding of a non-integer leads to a superposition of approximations.}
\end{figure*} 

%%%%%%%%%%%%%%%%%%%%%%%%%%%%%%%%%%%%%%%%%%%%%%%%%%%%%%%%%%%%%%%%%%%%%%%%%%%
\section*{Code Availability}
%%%%%%%%%%%%%%%%%%%%%%%%%%%%%%%%%%%%%%%%%%%%%%%%%%%%%%%%%%%%%%%%%%%%%%%%%%%
All code associated with this publication can be found in the IBM Qiskit Optimization module, currently hosted at \href{https://github.com/Qiskit/qiskit-optimization}{https://github.com/Qiskit/qiskit-optimization}.
 
%%%%%%%%%%%%%%%%%%%%%%%%%%%%%%%%%%%%%%%%%%%%%%%%%%%%%%%%%%%%%%%%%%%%%%%%%%%
\section*{Acknowledgments}
%%%%%%%%%%%%%%%%%%%%%%%%%%%%%%%%%%%%%%%%%%%%%%%%%%%%%%%%%%%%%%%%%%%%%%%%%%%
 
This material is for informational purposes only and is not the product of JPMorgan Chase \& Co.’s Research Department.  This material is not intended as research, a recommendation, advice, offer or solicitation for the purchase or sale of any financial product or service, and is not a research report and is not intended as such. This material is not intended to represent any position or opinion of JPMorgan Chase \& Co.  JPMorgan Chase \& Co. disclaims any responsibility or liability whatsoever for the quality, accuracy or completeness of the information herein, and for any reliance on, or use of this material in any way. \copyright 2021 JPMorgan Chase \& Co.
 
IBM, the IBM logo, and ibm.com are trademarks of International Business Machines Corp., registered in many jurisdictions worldwide. Other product and service names might be trademarks of IBM or other companies. The current list of IBM trademarks is available at \url{https://www.ibm.com/legal/copytrade}.

%%%%%%%%%%%%%%%%%%%%%%%%%%%%%%%%%%%%%%%%%%%%%%%%%%%%%%%%%%%%%%%%%%%%%%%%%%%
\appendix
%%%%%%%%%%%%%%%%%%%%%%%%%%%%%%%%%%%%%%%%%%%%%%%%%%%%%%%%%%%%%%%%%%%%%%%%%%%

%%%%%%%%%%%%%%%%%%%%%%%%%%%%%%%%%%%%%%%%%%%%%%%%%%%%%%%%%%%%%%%%%%%%%%%%%%%
\section{\label{sec:quantum_dictionary} Quantum Dictionary}
%%%%%%%%%%%%%%%%%%%%%%%%%%%%%%%%%%%%%%%%%%%%%%%%%%%%%%%%%%%%%%%%%%%%%%%%%%%
 
The \emph{Quantum Dictionary} was introduced in~\cite{Gonciulea2019} as a quantum computing pattern for encoding functions, in particular polynomials, into a quantum state using geometric sequences. The paper shows how quantum algorithms like search and counting applied to a quantum dictionary allow to solve combinatorial  optimization and QUBO problems more efficiently than using classical methods.

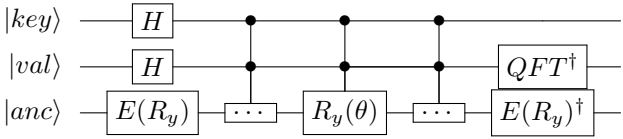
\begin{figure}[ht]
    \centering
    \mbox{
    \Qcircuit @C=1em @R=0em @!R {
    & \lstick{\ket{key}{}} & \gate{H} &\ctrl{2} & \ctrl{2} & \ctrl{2} & \qw & \qw  \\
    & \lstick{\ket{val}{}} & \gate{H} &\ctrl{1} & \ctrl{1} & \ctrl{1} \qw & \gate{QFT^\dag}  & \qw   \\ 
    & \lstick{\ket{anc}{}} & \gate{E(R_y)} & \gate{\text{\ldots}} & \gate{R_y(\theta)} & \gate{\text{\ldots}} & \gate{E(R_y)^\dag} & \qw   \\ 
    }
}
\caption{\label{fig:encoding-initial-state}Quantum Dictionary circuit.}
\end{figure} 

Encoding a geometric sequence can be done using the phase gate as we explained earlier, but it is worth mentioning an alternative method described in~\cite{Gonciulea2019}, which uses the $R_y$ family of gates.

\begin{figure}[ht]
    \centering
    \mbox{
    \Qcircuit @C=1em @R=1em {
    & \lstick{\ket{0}{}} & \gate{R_x(\pi/2)} & \gate{Z} & \gate{X} & \qw & \push{\rule{.3em}{0em}=\rule{.3em}{0em}} & \gate{E(R_y)} & \qw \\ 
    }
}
\caption{\label{fig:eigenstate-preparation}Eigenstate preparation for $R_y$.}
\end{figure} 
 
The $R_y$ gate is applied to an ancillary register containing one of its eigenstates prepared by an operator $E(R_y)$, conditioned on both the key and value registers. The rotation angle $\theta$ is different for each application, representing a number that contributes to the values corresponding to keys that have the conditioned key as a subset. In particular, when encoding a polynomial, a rotation will be applied for each of its coefficients.
 
In~\cite{Gonciulea2019} we used the $R_y$ operator, with the eigenstate $1/\sqrt{2}(i\ket{0} + \ket{1})$, independent of the rotation angle. This state can be prepared by the circuit in Fig.~\ref{fig:eigenstate-preparation}. The corresponding eigenvalue of $R_y(2\theta)$ is $e^{i\theta}$.

%%%%%%%%%%%%%%%%%%%%%%%%%%%%%%%%%%%%%%%%%%%%%%%%%%%%%%%%%%%%%%%%%%%%%%%%%%%
\section{\label{sec:handling-non-integers}Handling Non-Integers}
%%%%%%%%%%%%%%%%%%%%%%%%%%%%%%%%%%%%%%%%%%%%%%%%%%%%%%%%%%%%%%%%%%%%%%%%%%%

When handling non-integers, we have two choices.
The first is to approximate them with fractions with a common denominator and encode the numerator into quantum registers before performing computations, cf.~Appendix \ref{sec:approx-coefficients}.
The second, cf.~Appendix \ref{sec:encoding-non-integers}, is to phase-encode real numbers and let the inverse QFT convert the result of the computation into a superposition of approximations.

%%%%%%%%%%%%%%%%%%%%%%%%%%%%%%%%%%%%%%%%%%%%%%%%%%%%%%%%%%%%%%%%%%%%%%%%%%%
\subsection{\label{sec:approx-coefficients}Approximating Real Coefficients by Fractions}
%%%%%%%%%%%%%%%%%%%%%%%%%%%%%%%%%%%%%%%%%%%%%%%%%%%%%%%%%%%%%%%%%%%%%%%%%%%
 
If we relax the assumption that all coefficients are integers, we can approximate non-integers by dividing all values in $\mu$ and $\Sigma$ by the largest (scaling the range of the coefficients to $[-1, 1)$), and approximating each value $k$ by a fraction $\frac{k}{2^m}$ with $-2^{m - 1} \le k < 2^{m - 1}$, where $m$ is the number of value qubits.
As an example, suppose $n=3$, $m=5$, $q=0.5$, and
\begin{eqnarray}
\fontsize{8pt}{12pt}\selectfont
\mu = \left(  
\begin{array}{c}
\num{-3.77e-3} \\
\num{1.09e-3} \\
\num{2.41e-3}
\end{array}
\right).
\end{eqnarray}
%\begin{eqnarray}
%\fontsize{8pt}{12pt}\selectfont
%\Sigma = \left(  
%\begin{array}{ccc}
%\num{4.96e-04} & \num{-1.98e-04} & \num{5.60e-05} \\
%\num{-1.98e-04} & \num{2.70e-04} & \num{6.26e-03} \\
%\num{5.60e-05} & \num{6.26e-03} & \num{2.54e-04}
%\end{array}
%\right)
%\end{eqnarray}
Scaling the coefficients of $\mu$ leads to
\begin{eqnarray}
\fontsize{8pt}{12pt}\selectfont
\mu = \left(  
\begin{array}{c}
\num{-1.0} \\
\num{0.29} \\
\num{0.64}
\end{array}
\right),
\end{eqnarray}
and approximating them by fractions leads to
\begin{eqnarray}
\mu = \left(  
\begin{array}{c}
-16/32 \\
5/32 \\
10/32
\end{array}
\right).
\end{eqnarray}
As the approximated function coefficients have a common denominator, we can ignore the denominator and treat the values as integers
\begin{eqnarray}
\fontsize{8pt}{12pt}\selectfont
\mu = \left(  
\begin{array}{c}
-16 \\
5 \\
10
\end{array}
\right),
\end{eqnarray}
which results in the optimization problem
\begin{eqnarray}
\min_{x \in \{0, 1\}^3} -(-16x_1 + 5x_2 + 10x_3).
\end{eqnarray}

%%%%%%%%%%%%%%%%%%%%%%%%%%%%%%%%%%%%%%%%%%%%%%%%%%%%
\subsection{\label{sec:encoding-non-integers}Encoding Real Coefficients as Fej\'{e}r Distributions}
%%%%%%%%%%%%%%%%%%%%%%%%%%%%%%%%%%%%%%%%%%%%%%%%%%%%

Recall the unitary operator $U_G(\theta)$ from Fig.~\ref{fig:ug-definition}, where $\theta \in [-\pi, \pi)$.
As discussed in Sec.~\ref{subsec:construction_of_a}, the process for encoding an integer $-2^{m - 1} \le k < 2^{m - 1}$ is to apply $U_{\text{G}}(\frac{2\pi}{2^{m}}k)$ to an $m$-qubit register in equal superposition, followed by a single application of the inverse Quantum Fourier Transform.

\begin{figure}[ht]
\centering
    \mbox{
    \Qcircuit @C=0.7em @R=1em {
    \push{H^{\otimes m}\ket{0}_m{}} & & \gate{U_{\text{G}}(\theta)}  & \gate{QFT^\dag} & \qw &
    \push{=} & 
    \push{U_{\text{Fej\'{e}r}}(\theta) \ket{0}_m}
    }
    
    }
\caption{\label{fig:geom-qft-angle}$U_{\text{G}}(\theta)$ followed by inverse $QFT$.}
\end{figure}

For the general case of a real number, shown in Fig.~\ref{fig:geom-qft-angle}, where angle $\theta \in [-\pi, \pi)$.
The application of the same sequence of gates from the integer case results in a quantum state whose state vector consists of the inner product between $G(\theta)$ and the Fourier bases $G(\frac{2\pi}{2^{m}}j)$, representing a similarity measure between $\theta$ and $\frac{2\pi}{2^{m}}j$, for $0 \le j \le 2^m - 1$, where $G(\theta)$ denotes the geometric sequence vector $(1, e^{i\theta}, \ldots, e^{i(2^m - 1)\theta})$.
We will call the operator preparing this state $U_{\text{Fej\'{e}r}}$ because the outcome probability distribution is the Fej\'{e}r distribution~\cite{Janson2010}:

\begin{eqnarray}
U_{\text{Fej\'{e}r}}(\theta) \ket{0}_m = \frac{1}{\sqrt{2^m}}\sum_{j=0}^{2^m - 1} \langle G(\theta) , G(\frac{2\pi}{2^{m}}j)\rangle \ket{j}
\end{eqnarray}

If $\theta=\frac{2\pi}{2^{m}}a$, for a real number $-2^{m-1} \le a < 2^{m-1}$, then $U_{\text{Fej\'{e}r}}(\theta)\ket{0}_m$ prepares a state whose two most likely measurement outcomes are the closest two integers to $a$.
The probability of measuring one of them is at least 81\% ~\cite{Nielsen2011}.
Fig.~\ref{fig:fejer-example} shows the probability distribution of the outcomes for multiple values of $a$ where $m=4$.

%%%%%%%%%%%%%%%%%%%%%%%%%%%%%%%%%%%%%%%%%%%%%%%%%%%%%%%%%%%%%%%%%%%%%%%%%%%
\section{\label{sec:hardware-config}Hardware Specifications}
%%%%%%%%%%%%%%%%%%%%%%%%%%%%%%%%%%%%%%%%%%%%%%%%%%%%%%%%%%%%%%%%%%%%%%%%%%%

\begin{figure}[ht]
\centering
\includegraphics[width=7cm]{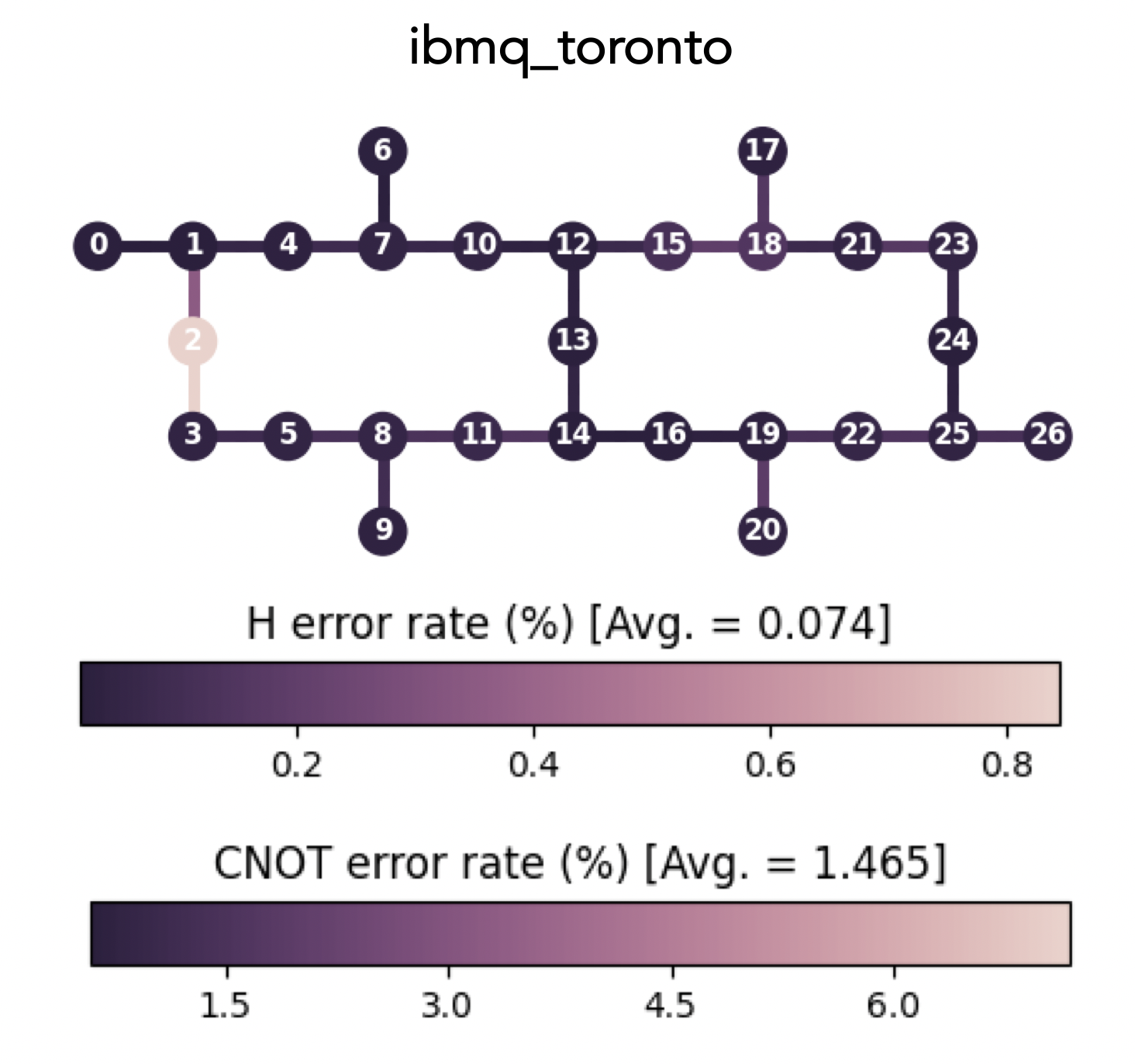}
\caption{\label{fig:harware-specs}The error map for ibmq\_toronto.}
\end{figure}

The error map for ibmq\_toronto in Fig.~\ref{fig:harware-specs} was generated on the day of experimentation using Qiskit's visualization library~\cite{qiskit}.

%%%%%%%%%%%%%%%%%%%%%%%%%%%%%%%%%%%%%%%%%%%%%%%%%%%%%%%%%%%%%%%%%%%%%%%%%%%
\bibliographystyle{unsrtnat}
\bibliography{main}% Produces the bibliography via BibTeX.
%%%%%%%%%%%%%%%%%%%%%%%%%%%%%%%%%%%%%%%%%%%%%%%%%%%%%%%%%%%%%%%%%%%%%%%%%%%
 
\end{document}